\documentclass[aps,prd,preprintnumbers,showpacs,showkeys,nofootinbib,
superscriptaddress,fleqn,floatfix,tightenlines,10pt]{revtex4-1}
\usepackage{amsmath,amsfonts,amssymb,amscd,amsxtra,amsthm}
\usepackage{graphicx}  
\usepackage{epstopdf}
\usepackage{dcolumn}  
\usepackage{bm}          
\usepackage{slashed}
\usepackage{cancel}
\usepackage{float}
\usepackage{mathtools}
\usepackage{amsbsy}
\usepackage{amstext}

\usepackage[utf8]{inputenc} 
\usepackage{booktabs}
\usepackage[normalem]{ulem} 
\usepackage[dvipsnames]{xcolor} 
\usepackage{tabularx}
\usepackage{enumitem}  
\usepackage{array} 
\usepackage{slashed}
\usepackage{tikz}
\usepackage{float}
\usepackage{multirow}
\renewcommand\sout{\bgroup \color{red} \ULdepth=-.5ex \ULset}

\makeatletter

\begin{document}
\preprint{INHA-NTG-11/2018}
\title{Effects of nucleon resonances on $\eta$ photoproduction off the 
  neutron reexamined} 
\author{Jung-Min Suh}
\email[E-mail: ]{suhjungmin@inha.edu}
\affiliation{Department of Physics, Inha University, Incheon 22212,
 Republic of Korea}
\author{Sang-Ho Kim}
\email[E-mail: ]{sangho\_kim@korea.ac.kr}
\affiliation{Center for Extreme Nuclear Matters (CENuM),
Korea University, Seoul 02841, Republic of Korea}
\affiliation{Department of Physics, Pukyong National University
  (PKNU), Busan 48513, Republic of Korea}
\author{Hyun-Chul Kim}
\email[E-mail: ]{hchkim@inha.ac.kr}
\affiliation{Department of Physics, Inha University, Incheon 22212,
 Republic of Korea}
\affiliation{Advanced Science Research Center, Japan Atomic Energy
  Agency, Shirakata, Tokai, Ibaraki, 319-1195, Japan} 
\affiliation{School of Physics, Korea Institute for Advanced Study 
 (KIAS), Seoul 02455, Republic of Korea}
\date{\today}
\begin{abstract}
We investigate $\eta$ photoproduction off the neutron target, i.e.,
$\gamma n \to \eta n$, employing an effective Lagrangian method combining 
with a Regge approach.
As a background, we consider nucleon exchange in the $s$-channel diagram 
and $\rho$- and $\omega$-meson Regge trajectories in the $t$ channel.
The role of nucleon resonances given in the Review of Particle Data Group 
in the range of $W \approx 1500 - 2100$ MeV and the narrow nucleon 
resonance $N(1685,1/2^+)$ is extensively studied.
The numerical results of the total and differential cross sections,
double polarization observable $E$, and helicity-dependent cross 
sections $\sigma_{1/2}$, $\sigma_{3/2}$ are found to be in qualitative
agreement with the recent A2 experimental data.
The predictions of the beam asymmetry are also given.
\end{abstract}

\maketitle

\section{Introduction}
\label{Sec:1}
Photoproduction of mesons provides an essential tool to investigate
excited baryons. In particular,  $\eta$ photoproduction plays a role
of a filter for excited nucleon resonances, since $\eta$ is a
pesudoscalar and isoscalar meson, and it contains hidden strangeness. 
So, only the selected number of the excited nucleon resonances can be 
exclusively studied, which can be only coupled to the $\eta$
meson~\cite{Tanabashi:2018}. Since Kuznetsov et
al.~\cite{Kuznetsov:2006kt} reported the evidence of the narrow
bump-like structure around the center-of-mass (CM) energy $W\sim 1.68$ GeV
in $\eta$ photoproduction off the quasi-free neutron, 
a number of experiments on $\eta$ photoproduction off the neutron has
been performed by the Tohoku group at the Laboratory of Nuclear
Science (LNS) at Tohoku University~\cite{Miyahara:2007zz}, 
CBELSA/TAPS Collaboration~\cite{Jaegle:2008ux, Jaegle:2011sw,
  Witthauer:2017pcy}, and A2 Collaboration~\cite{Werthmuller:2013rba,
  Witthauer:2013tkm, Werthmuller:2014thb}. Starting from the $\eta N$
threshold, the total cross section of $\eta$ photoproduction off the
nucleon raises rapidly because of the dominant and broad excited
nucleon resonance $N(1535,1/2^-)$. It falls off somewhat slowly
from $W\sim 1535\,\mathrm{MeV}$. The narrow bump-like
structure is then observed on the shoulder of the $N(1535,1/2^-)$ in
the vicinity of $W\sim 1.68$ GeV but only in $\eta$ photoproduction
off the neutron. Such a narrow structure was not found in the proton
target as reported by Refs.~\cite{Bartalini:2007fg, Jaegle:2008ux} or
a dip-like structure appears~\cite{McNicoll:2010qk,
  Jaegle:2011sw}. The finding that the narrow bump-like
structure is only clearly seen in $\eta$ photoproduction off the
neutron is coined \textit{neutron anomaly}~\cite{Kuznetsov:2008hj}.

Theoretical interpretations on this narrow bump-like structure are not
in consensus, whereas the evidence of its existence has been firmly 
established.  Right after the finding of it,
Refs.~\cite{Choi:2005ki,Choi:2007gy, Fix:2007st}
explained the $\gamma n\to \eta n$ reaction very well within an
Effective Lagrangian approach, regarding the narrow structure as the
narrow nucleon resonance $N(1685,1/2^+)$. These works were motivated
by the results from the chiral quark-soliton model
($\chi$QSM)~\cite{Polyakov:2003dx, Kim:2005gz}.
We want to mention that recently, A2 Collaboration has carried out the
measurement of the double polarization observables and the
helicity-dependent cross sections for $\eta$ photoproduction from the
quasi-free proton and neutron~\cite{Witthauer:2017get,Witthauer:2017wdb}. 
Interestingly, the narrow bump-like structure was seen only in the 
spin-$1/2$ helicity-dependent cross section. A narrow
structure is experimentally favored to be interpreted as a narrow
$P_{11}$ nucleon resonance as mentioned by
Ref.~\cite{Witthauer:2017get}, which supports the analyses of
Refs.~\cite{Choi:2005ki,Choi:2007gy, Fix:2007st}. On the other hand,  
there have been various theoretical disputes on the interpretation of
the narrow bump-like structure as a narrow $N(1685,1/2^+)$ resonance.  
Ref.~\cite{Shklyar:2006xw} proposed as the nature of the narrow
structure the coupled-channel effects by $N(1650,1/2^-)$ and
$N(1710,1/2^+)$ within the unitary coupled-channels effective
Lagrangian approach. Reference~\cite{Shyam:2008fr}
considered it as the interference effects of $N(1535,1/2^-)$,
$N(1650,1/2^-)$, $N(1710,1/2^+)$, and $N(1720,3/2^+)$ resonance 
contributions, based on a coupled-channels $K$-matrix 
method. The Bonn-Gatchina partial-wave analysis (Bn-Ga PWA)
group~\cite{Anisovich:2008wd, Anisovich:2015tla, Anisovich:2017xqg}  
regarded the narrow bump-like structure as the effects arising from 
interference between $N(1535,1/2^-)$ and $N(1650,1/2^-)$.
The Bn-Ga PWA group argued that the inclusion of the narrow nucleon
resonance made the results worse within their PWA approach.
On the other hand, D\"oring and Nakayama~\cite{Doring:2009qr}
examined the ratio of the cross section $\sigma_n/\sigma_p$ with the
intermediate meson-baryon states with strangeness and considered the
narrow structure as the effects coming from the opening of
the strangeness channel in intermediate states.  
Kuznetsov et al.~\cite{Kuznetsov:2008hj, Kuznetsov:2017qmo,
  Kuznetsov:2017ayk} rebutted that interpretation of the narrow
bump-like structure as an interference effect between the $S$-wave
nucleon resonances. Moreover, both the narrow excited proton and
neutron were also seen in Compton scattering $\gamma N\to \gamma
N$~\cite{Kuznetsov:2010as, Kuznetsov:2015nla} and the reactions  
$\gamma N\to \pi\eta N$~\cite{Kuznetsov:2017xgu}. 

Meanwhile, two of the authors recently studied $K^0\Lambda$
phopoproduction off the neutron based on an effective Lagrangian
approach combined with a Regge model~\cite{Kim:2018qfu}. 
They found that the narrow resonance $N(1685,1/2^+)$ comes into critical 
play to describe the experimental data on the differential cross sections.
The corresponding CLAS data show a dip structure in forward angle regions 
at the pole position of the narrow nucleon resonance, although the signs 
are not clear since the width of the dip is not as large as the energy bin
of the data~\cite{Compton:2017xkt}. 
On the other hand, a different conclusion
was drawn by the Bn-Ga PWA group~\cite{Anisovich:2017afs} with the
same CLAS data, in which the evidence of the $N(1685,1/2^+)$ was
discarded. Note that this dip structure was not observed
experimentally in $K^+\Lambda$ photoproduction, although there are
relevant theoretical studies to support its
existence~\cite{Mart:2011ey,Mart:2013fia}. 

In view of this puzzled situation related to the existence of
$N(1685,1/2^+)$, we want to reexamine $\eta$ photoproduction off the
neutron within the framework of the effective Lagrangian approach
combined with a Regge model, focusing on the role of $N(1685,1/2^+)$.
Since the previous works~\cite{Choi:2005ki,Choi:2007gy} were performed
by using old experimental information on the excited nucleon
resonances and more experimental data on the narrow $N(1685,1/2^+)$
resonances were compiled as explained above, it is worthwhile to
reinvestigate the roles of the excited nucleon resonances in $\eta$
photoproduction off the neutron with $N(1685,1/2^+)$ also included as a
nucleon resonance. We include altogether sixteen different excited
nucleon resonances in the $s$ channel, fixing all relevant parameters
by using the experimental data and empirical information. Each
contribution of the excited nucleons will be scrutinized. The $t$
channel will be described by vector Reggeon exchanges. 

This paper is organized as follows: In Section~\ref{SecII} we explain
the general formalism of the effective Lagrangian approach together
with the Regge model in detail. In Section~\ref{SecIII}, we present
the numerical results of the total and differential cross sections, and
the double polarization observables for the $\gamma n \to \eta n$
reaction and discuss physical implications of them. Secion~\ref{SecIV}
is devoted to summary and outlook.

\section{General formalism}
\label{SecII}
\begin{figure}[htp]
\centering
\includegraphics[width=11cm]{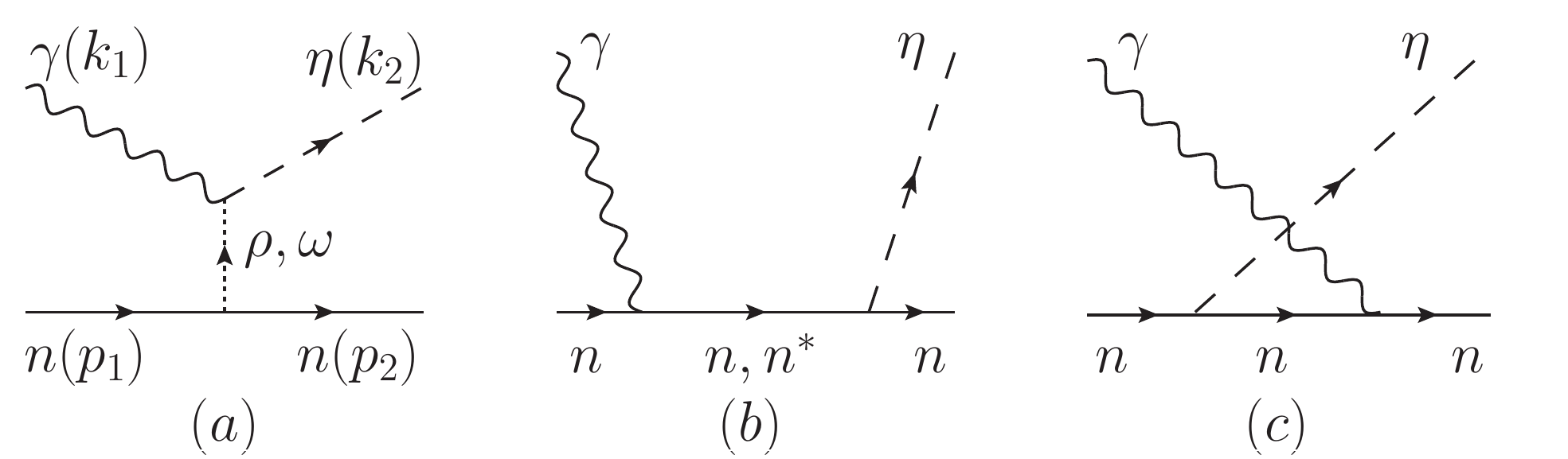}
\caption{Feynman diagrams for the $\gamma n \to \eta n$ reaction.}
\label{fig:1}
\end{figure}

In this paper, we study $\eta$ photoproduction off the neutron in an
effective Lagrangian approach with a Regge method. 
The scattering amplitudes for this process can be divided into two 
parts, i.e., the background and excited nucleon resonance ($N^*$)
contributions. Firstly, we consider $\rho$ and $\omega$ Reggeon
exchanges in the $t$ channel as the background contribution that are
shown in Fig.~\ref{fig:1}(a) in which the symbols in parentheses stand
for the four-momenta of the corresponding paticles. These two
vector-Reggeon exchanges are enough to describe the $\gamma n
\to \eta n$ reaction at higher energies. $N$ exchanges in both the $s$
(Fig.~\ref{fig:1}(b)) and $u$ (Fig.~\ref{fig:1}(c)) channels are also
taken into account as the background contributions, although their
effects are tiny on the cross sections. Secondly, the $N^*$
contributions are included in the $s$-channel diagram in addition to
nucleon exchange (Fig.~\ref{fig:1}(b)), of which information is taken
from the Review of Particle Physics~\cite{Tanabashi:2018}.
In the present work, we introduce fifteen nucleon resonances that
are coupled strongly to the $\gamma N$ and $\eta N$ vertices.
We regard the narrow bump-like structure as the narrow resonance
$N(1685,1/2^+)$ of which the existence was predicted in the
$\chi$QSM~\cite{Polyakov:2003dx, Kim:2005gz}.    

We start with the background contributions. The photon vertices can be
constructed by using the following effective Lagrangians
\begin{align}
\mathcal L_{\gamma \eta V} &=
\frac{e g_{\gamma \eta V}}{4M_\eta} \epsilon^{\mu\nu\alpha\beta}
F_{\mu\nu} V_{\alpha\beta} \phi_\eta + \mathrm{H.c.},
\cr
\mathcal L_{\gamma NN} &=
- \bar N \left[ e_N \gamma_\mu - \frac{e\kappa_N}{2M_N}
\sigma_{\mu\nu}\partial^\nu \right] A^\mu N,
\label{eq:BornLag1}
\end{align}
where $F_{\mu\nu}$ and $V_{\alpha\beta}$ denotes the electromagnetic
and vector-meson field-strength tensors defined respectively by 
$F_{\mu\nu} = \partial_\mu A_\nu - \partial_\nu A_\mu$ and
$V_{\alpha\beta} = \partial_\alpha V_\beta - \partial_\beta
V_\alpha$. $A_\mu$, $\phi_\eta$, and $N$ stand for the photon,
pseudoscalar $\eta$ meson, and nucleon fields, respectively.
$V$ represents generically either the $\rho$- or $\omega$-vector
mesons. $M_{h}$ is the mass of the hadron $h$ involved in the
process. $e_N$ designates the electric charge and $e$ the unit
electric charge. In the present work, we need to consider only the
magnetic part in the $\gamma NN$ vertex  because the charge of the
$\eta$ meson is neutral.
The coupling constant $g_{\gamma \eta V}$ is extracted from the
experimental data on the corresponding decay width of the vector meson
$V$ 
\begin{align}
\Gamma_{V \to \eta \gamma} = \frac{g_{\gamma \eta V}^2}{12\pi}
\frac{e^2}{M_\eta^2} \left( \frac{M_V^2-M_\eta^2}{2M_V} \right)^3.
\label{eq:DW:VEtaG}
\end{align}
Since the decay widths of the $\rho$ and $\omega$ mesons are
experimentally known to be $\Gamma_{\rho \to \eta \gamma} = 45.5$ keV and
$\Gamma_{\omega \to \eta \gamma} = 3.82$ keV~\cite{Tanabashi:2018},
one can easily determine the vector coupling constants as follows
\begin{align}
g_{\rho\eta\gamma} = 0.91, \,\,\, 
g_{\omega\eta\gamma} = 0.24.
\label{eq:Coupl:VEtaG}
\end{align}
The anomalous magnetic moment of the neutron is also given by the
PDG~\cite{Tanabashi:2018}: $\kappa_n =-1.91$. 

The effective Lagrangians for the meson-nucleon interactions are
written as
\begin{align}
\mathcal L_{VNN} &=
-g_{VNN} \bar N \left[ \gamma_\mu N - 
\frac{\kappa_{VNN}}{2M_N} \sigma_{\mu\nu} N
\partial^\nu \right] V^\mu + \mathrm{H.c.},
\cr
\mathcal L_{\eta N N} &=
\frac{g_{\eta N N}}{2 M_N} \bar N \gamma_\mu \gamma_5 N  \partial^\mu 
\phi_\eta.
\label{eq:BornLag2}
\end{align}
The relevant strong coupling constants are taken from the Nijmegen
soft-core  potential~\cite{Stoks:1999bz}
\begin{align}
g_{\rho N N} = 2.97, \,\,\, \kappa_{\rho NN} = 4.22, \,\,\,
g_{\omega N N} = 10.4, \,\,\, \kappa_{\omega NN} = 0.41, \,\,\,
g_{\eta N N} = 6.34.
\label{eq:Coupl:NP}
\end{align}

In order to construct the invariant amplitude for vector-Reggeon
exchange,  We follow the Regge formalism of
Refs.~\cite{Donachie2002,Guidal:1997hy}. 
In general, the Regge amplitude is obtained by replacing the Feynman 
propagators by the Regge propagators as follows 
\begin{align}
\frac{1}{t-M_V^2} \to P_V^{\mathrm{Regge}} (t) =
\left( \frac{s}{s_0} \right)^{\alpha_V(t)-1}
\frac{1}{\sin[\pi\alpha_V(t)]}
\frac{\pi\alpha_V'}{\Gamma[\alpha_V(t)]} D_V(t).
\label{eq:REGGEPRO}
\end{align}
We fix the energy-scale parameter to be $s_0 = 1\,\mathrm{GeV^2}$ for
simplicity.
To determine the signature factor $D_V(t)$, we refer to pion
photoproduction $\gamma p \to \pi^0 p$~\cite{Guidal:1997hy}, where the
same Regge trajectories are considered in the $t$-channel.
It is demonstrated that the $\rho$-meson trajectory should be
degenerate to describe the proper asymptotic high-energy behavior
whereas the $\omega$ trajectory is nondegenerate to explain the dip
structure for the  $d\sigma/dt$ at a certain point.
The signature factors are explicitly expressed as 
\begin{align}
D_\rho(t) = \mathrm{exp}(-i\pi\alpha_\rho(t)),\,\,\,
D_\omega(t) = \frac{\mathrm{exp}(-i\pi\alpha_\omega(t))-1}{2} ,
\label{eq:SigFac}
\end{align}
and the vector-meson Regge trajectories read~\cite{Guidal:1997hy}
\begin{align}
\alpha_\rho (t) =  0.55 + 0.8t, \,\,\, \alpha_\omega (t) = 0.44 + 0.9t.
\label{eq:ReggeTraj}
\end{align}
Note that the invariant amplitudes for these background parts are all 
separately gauge invariant by construction. 

To respect the finite sizes of hadrons involved, we need to introduce
the empirical form factors. We mention that the form
factors are in effect main sources for model uncertainties. We use the
following type of the form factors for $N$ exchange in the $s$
channel: 
\begin{align}
F_N(q^2) = \frac{\Lambda^4} {\Lambda^4+\left(q^2-M_N^2\right)^2},
\label{eq:FF1}
\end{align}
where $q^2$ is the squared momentum of $q_s=k_1+p_1$ or $q_u=p_2-k_1$.
In the case of the $t$-channel Reggeon exchange, we introduce a scale 
factor which plays a role of the form factor:
\begin{align}
C_V(t) = \frac{a}{(1 - t / \Lambda^2)^2}.
\label{eq:ScaleFac}
\end{align}
In the present work, we use the fixed values of the cut-off masses
$\Lambda_N = 1.0$ GeV to avoid
additional ambiguities. The scaling parameters are taken to be 
$a_{\rho,\omega} = 1.4$ and $\Lambda_{\rho,\omega} = 1.0$ GeV.

There are many excited nucleon resonances above the $\eta N$
threshold ~\cite{Tanabashi:2018}, among which we select fifteen
$N^*$s in the range of $W \approx  (1500-2100)$ MeV as 
listed in Table~\ref{TAB1}. Note that $N(1990,7/2^+)$,
$N(2040,3/2^+)$, $N(2060,5/2^-)$ and $N(2100,1/2^+)$ are excluded
because of lack of information as to how they are coupled to the
photon or the $\eta$ meson. In addition to the fifteen nucleon
resonances, we consider the narrow resonance $N(1685,1/2^+)$.  
The effective Lagrangians for the photo-excitations $\gamma N \to
N^*$ are given in the following forms:  
\begin{align}
\mathcal{L}^{1/2^\pm}_{\gamma  N N^*} &= 
\frac{eh_1}{2M_N} \bar N \Gamma^\mp
\sigma_{\mu\nu} \partial^\nu A^\mu N^* + \mathrm{H.c.} ,               
\cr
\mathcal{L}^{3/2^\pm}_{\gamma N N^*}&= 
-ie \left[ \frac{h_1}{2M_N} \bar N \Gamma_\nu^\pm
 - \frac{ih_2}{(2M_N)^2} \partial_\nu \bar N
 \Gamma^\pm \right] F^{\mu\nu} N^*_\mu + \mathrm{H.c.},   
\cr
\mathcal{L}^{5/2^\pm}_{\gamma N N^*} &=
e\left[ \frac{h_{1}}{(2M_N)^2} \bar N \Gamma_\nu^\mp
-\frac{ih_{2}}{(2M_N)^3} \partial_\nu \bar N
\Gamma^\mp \right] \partial^\alpha F^{\mu\nu}
N^*_{\mu\alpha} + \mathrm{H.c.} ,
\cr
\mathcal{L}^{7/2^\pm}_{\gamma N N^*} &=
ie \left[ \frac{h_{1}}{(2M_N)^3} \bar N \Gamma_\nu^\pm
-\frac{ih_{2}}{(2M_N)^4} \partial_\nu \bar N
\Gamma^\pm \right] \partial^\alpha \partial^\beta F^{\mu\nu}
N^*_{\mu\alpha\beta} + \mathrm{H.c.} ,
\label{eq:ResLag1}
\end{align}
where the superscripts denote the spin and parity of the corresponding
nucleon resonances. $N^*$ stands for the spin-1/2 excited nucleon
field whereas $N^*_\mu$, $N^*_{\mu\alpha}$, and $N^*_{\mu\alpha\beta}$
represent the Rarita-Schwinger fields of spin-3/2, -5/2, and -7/2,
respectively. $\Gamma^\pm$ and $\Gamma_\nu^\pm$, which are related to
the parity of an excited nucleon involved, are defined by  
\begin{align}
\Gamma^{\pm} = \left(
\begin{array}{c}
\gamma_5 \\ I_{4\times4}
\end{array} \right) ,
\,\,\,\,
\Gamma_\nu^{\pm} = \left(
\begin{array}{c}
\gamma_\nu \gamma_5 \\ \gamma_\nu
\end{array} \right) .
\label{eq:GammaPM}
\end{align}
To determine the magnetic transition moments $h_{1,2}$, we relate
them to the Breit-Wigner helicity amplitudes
$A_{1/2,3/2}$~\cite{Oh:2007jd, Oh:2011}. The $A_{1/2,3/2}$ are taken
from the PDG~\cite{Tanabashi:2018}. Some ambiguities are
contained in $A_{1/2,3/2}$. In the present work, we choose the central
values of them. Although we could obtain better theoretical results
by fitting these couplings, we would not perform it because the
purpose of this work is to investigate how far we can describe the
experimental data when the narrow $N^*(1685,1/2^+)$ is explicitly
included. Thus, in the procedure, it is essential to reduce any
theoretical ambiguities such that the contribution of this narrow
nucleon resonance can be carefully examined.  
All the relevant the helicity amplitudes and the photo-coupling
constants are summarized in Table~\ref{TAB1}. 
The magnetic transition moment of the narrow resonance $N(1685,1/2^+)$
is taken from Ref.~\cite{Yang:2018gju}.
\begin{table}[h]
\caption{The fifteen nucleon resonances taken from the
PDG~\cite{Tanabashi:2018} and the numerical values of the magnetic
transition moments. The helicity amplitudes $A_{1/2,\,3/2}$ are given in
units of $10^{-3}/\sqrt{\mathrm{GeV}}$. In addition, we include the
narrow nucleon resonance $N(1685,1/2^+)$.} 
\label{TAB1}
\begin{tabular}{c||cc|cccc}
\hline
State&Rating&Width [MeV]
&$A_{1/2}$&$A_{3/2}$&$h_1$ &$h_2$ \\
\hline
$N(1520,3/2^-)$&****&100-120 (110)
&$\approx -50$&$\approx -115$&$-0.77$&$-0.62$ \\
$N(1535,1/2^-)$&****&125-175 (140)
&$\approx -75$&$\cdots$&$-0.53 $&$\cdots$ \\
$N(1650,1/2^-)$&****&100-150 (125)
& $\approx -10$&$\cdots$
& $0.063$ &$\cdots$ \\
$N(1675,5/2^-)$&****&130-160 (145)
&$-60 \pm 5 $&$-85 \pm 10$&$4.88$&$5.45$ \\
$N(1680,5/2^+)$&****&100-135 (120)
& $\approx 30$&$\approx -35$ 
&$\cdots$ & $\cdots$ \\
\hline
$N(1700,3/2^-)$&***&100-300 (200)
&$25 \pm 10$&$-32 \pm 18$&$-1.43$&$1.64$ \\
$N(1710,1/2^+)$&****&80-200 (140)
&$-40 \pm 20$&$\cdots$&$0.24$&$\cdots$ \\
$N(1720,3/2^+)$&****&150-400 (250)
&$-80\pm 50$&$-140 \pm 65$&$1.50$&$1.61$ \\
$N(1860,5/2^+)$&**&300
&$21 \pm 13$&$34 \pm 17$&$0.28$&$1.09$ \\
$N(1875,3/2^-)$&***&120-250 (200)
&$10 \pm 6$&$-20 \pm 15$&$-0.55$&$0.54$ \\
\hline
$N(1880,1/2^+)$&***&200-400 (300)
&$-60 \pm 50$&$\cdots$&$0.30$&$\cdots$ \\
$N(1895,1/2^-)$&****&80-200 (120)
&$13 \pm 6$&$\cdots$&$0.067$&$\cdots$ \\
$N(1900,3/2^+)$&****&100-320 (200)
&$0\pm 30$&$-60 \pm 45$&$0.29$&$-0.56$ \\
$N(2000,5/2^+)$&**&300
&$-18 \pm 12$&$-35 \pm 20$&$-0.47$&$-0.56$ \\
$N(2120,3/2^-)$&***&260-360 (300)
&$110 \pm 45$&$40 \pm 30$&$-1.71$&$2.41$ \\
\hline
$N(1685,1/2^+)$&   &30
&$ $&$ $&$-0.315$~\cite{Yang:2018gju}&$ $ \\
\hline
\end{tabular}
\end{table}

The effective Lagrangians for the strong vertices $\eta N N^*$ are
expressed as 
\begin{align}
\mathcal{L}^{1/2^\pm}_{\eta N N^*}&= 
 - i g_{\eta N N^*} \phi_\eta \bar N 
 \Gamma^\pm N^* + \mathrm{H.c.},                                   
\cr
\mathcal{L}^{3/2^\pm}_{\eta N N^*}&=
 \frac{g_{\eta N N^*}}{M_\eta} \partial^\mu \phi_\eta \bar N
 \Gamma^\mp N^*_\mu + \mathrm{H.c.},
\cr
\mathcal{L}^{5/2^\pm}_{\eta N N^*} &=
 \frac{ig_{\eta N N^*}}{M_\eta^2} \partial^\mu \partial^\nu \phi_\eta
 \bar N \Gamma^\pm N^*_{\mu\nu} + \mathrm{H.c.},
\cr
\mathcal{L}^{7/2^\pm}_{\eta N N^*} &=
- \frac{g_{\eta N N^*}}{M_\eta^3} \partial^\mu \partial^\nu \partial^\alpha
\phi_\eta \bar N \Gamma^\mp N^*_{\mu\nu\alpha} + \mathrm{H.c.}.
\label{eq:ResLag2}
\end{align}
Though there are more terms in the Lagrangians for higher-spin nucleon
resonances, we do not need them, considering the angular momenta and
parity conservation of $\eta$ photoproduction. 
The magnitude of the strong coupling constants, $g_{\eta N N^*}$, can be 
extracted from the partial decay widths $\Gamma_{N^* \to \eta N}$ given by 
the experimental data of the PDG~\cite{Tanabashi:2018}.
However, their signs are unknown and, moreover, only the upper limits
are known for some of $\Gamma_{N^* \to \eta N}$. So, we use
information on these unknown decay amplitudes from the quark model 
predictions~\cite{Capstick:1998uh}. To do that, we employ the following 
relation~\cite{Kim:2017nxg,Kim:2018qfu}:
\begin{align}
\langle \eta(\vec{q})\,N(-\vec{q},m_f) | -i \mathcal{H}_\mathrm{int} |
N^*({\bf 0},m_j) \rangle
=4 \pi M_{N^*} \sqrt\frac{2}{|\vec{q}|} \sum_{\ell,m_\ell}
\langle \ell\, m_\ell\, {\textstyle\frac{1}{2}}\, m_f | j \,m_j \rangle 
Y_{\ell,m_\ell} ({\hat q}) G(\ell),
\label{eq:DA}
\end{align}
where $\langle\ell\,m_\ell\,\frac{1}{2}\,m_f|j \,m_j \rangle$ and 
$Y_{\ell,m_\ell} ({\hat q})$ are the Clebsch-Gordan coefficients and
spherical harmonics, respectively.
A partial decay width is then obtained from the decay amplitudes
$G(\ell)$
\begin{align}
\Gamma (N^* \to \eta N) = \sum_\ell |G(\ell)|^2 .
\label{eq:DA2}
\end{align}

The spin and parity of the nucleon resonance impose constraints on the 
relative orbital angular momentum $\ell$ of the $\eta N$ final state.
For example, in the cases of $j^P = \frac{1}{2}^+$ and $\frac{1}{2}^-$
resonances, only $p(\ell=1)$ and $s(\ell=0)$ waves are allowed,
respectively.
Finally, the relations between the decay amplitudes and the strong 
coupling constants for the vertices with $j^P =
(1/2^\pm,\,3/2^\pm,\,5/2^\pm, 7/2^\pm)$ nucleon resonances are derived
as follows:  
\begin{align}
G\left(\frac{1+P}{2}\right) &= \mp
\sqrt{\frac{|\vec{q}|(E_N \mp M_N)}{4\pi M_{N^*}}} 
g_{\eta N N^*}\,\,\,\,\mathrm{for}\,\,\,\,N^*(1/2^P),            
\cr
G\left(\frac{3-P}{2}\right) &= \pm
\sqrt{\frac{|\vec{q}|^3(E_N \pm M_N)}{12\pi M_{N^*}}}
\frac{g_{\eta N N^*}}{M_\eta}\,\,\,\,\mathrm{for}\,\,\,\,N^*(3/2^P), 
\cr
G\left(\frac{5+P}{2}\right)  &= \mp
\sqrt{\frac{|\vec{q}|^5(E_N \mp M_N)}{30\pi M_{N^*}}}
\frac{g_{\eta N N^*}}{M_\eta^2}\,\,\,\,\mathrm{for}\,\,\,\,N^*(5/2^P).
\cr
G\left(\frac{7-P}{2}\right)  &= \pm
\sqrt{\frac{|\vec{q}|^7(E_N \pm M_N)}{70\pi M_{N^*}}}
\frac{g_{\eta N N^*}}{M_\eta^3}\,\,\,\,\mathrm{for}\,\,\,\,N^*(7/2^P),
\label{eq:DA:1/2to7/1}
\end{align}
where the magnitude of the three-momentum and the energy for 
$N$ in the rest frame of the nucleon resonance are given respectively
as  
\begin{align}
|\vec{q}|= \frac{1}{2M_{N^*}}
\sqrt{[M_{N^*}^2 - (M_N + M_\eta)^2][M_{N^*}^2 - (M_N - M_\eta)^2]},
\,\,\,\,E_N = \sqrt{M_N^2 + |\vec{q}|^2}.
\label{eq:3-monen}
\end{align}

In Table~\ref{TAB2}, the relevant values for the strong decays are 
tabulated.
$\Gamma_{N^*}$ designates the decay width of $N^*$.
We use the values in parentheses in Table~\ref{TAB1} for them.
For the spin-$3/2$, -$5/2$, and -$7/2$ propagators, we employ the 
Rarita-Schwinger formalism~\cite{Berends:1979rv,Behrends:1957rup,
Chang:1967zzc,Rushbrooke:1966zz} as in Refs.~\cite{Oh:2007jd,
Oh:2011,Kim:2011rm,Kim:2012pz,Kim:2014hha}.
The off-shell terms of the Rarita-Schwinger fields are excluded.
The PDG data are only available for the three nucleon resonances above
1875 MeV, i.e., $N(1880,1/2^+)$, $N(1895,1/2^-)$, and $N(1900,3/2^+)$, 
we determine the signs of the corresponding strong coupling constants
phenomenologically. The numerical values of all the necessary coupling
constants $g_{\eta N N^*}$ are listed in the last column of Table~\ref{TAB2}.
\begin{table}[h]
\caption{The numerical values of the strong couplings of the nucleon  
resonances. The decay amplitudes $G(\ell)$ in units of MeV are
obtained from Ref.~\cite{Capstick:1998uh} and the branching ratios of
$N^*\to \eta N$ decays are taken from Ref.~\cite{Tanabashi:2018}.}
\label{TAB2}
\begin{tabular}{c||cc|cc|c}
\hline
State&$G(\ell)$
&$g_{\eta N N^*}$&$\Gamma_{N^* \to \eta N} / \Gamma_{N^*} [\%]$
&$|g_{\eta N N^*}|$&$g_{\eta N N^*}$(final) \\
\hline
$N(1520,3/2^-)$
&$0.4_{-0.4}^{+2.9}$&$-8.30$&$0.07-0.09$&$5.23- 6.49$&$-5.23$ \\
$N(1535,1/2^-)$
&$8.1 \pm 0.8$&$2.05$&$30-55$&$1.58-2.14$&$2.10$ \\
$N(1650,1/2^-)$
&$-2.4 \pm 1.6$&$-0.43$&$15-35$&$0.76-1.16$&$-0.80$ \\
$N(1675,5/2^-)$
&$-2.5 \pm 0.2$&$-2.50$&$< 1$&$<0.90 $&$-0.90$ \\
$N(1680,5/2^+)$
&$0.6 \pm 0.1$&$-2.98$&$< 1$&$<4.07$&$-2.47$ \\
\hline
$N(1700,3/2^-)$
&$-0.2 \pm 0.1$&$0.38$&seen& & $0.38$ \\
$N(1710,1/2^+)$
&$5.7 \pm 0.3$&$-4.23$&$10-50$&$2.93- 6.55$&$-4.00$ \\
$N(1720,3/2^+)$
&$5.7 \pm 0.3$&$2.08$&$1 -5$&$0.43-4.50$&$1.00$ \\
$N(1860,5/2^+)$
&$1.9 \pm 0.8$&$-2.84$&$2 -6$&$2.47- 4.27$&$-2.47$ \\
$N(1875,3/2^-)$
&$4.0 \pm 0.2$&$-3.58$&$< 1$&$<0.89 $&$-0.80$ \\
\hline
$N(1880,1/2^+)$
&$ $&$ $&$5 - 55$&$2.02-6.69$&$2.00$ \\
$N(1895,1/2^-)$
&$ $&$ $&$15 - 40$&$0.60-0.99$&$0.60$ \\
$N(1900,3/2^+)$
&$ $&$ $&$2 - 14 $&$0.33-0.87$&$0.33$ \\
$N(2000,5/2^+)$
&$1.9 \pm 0.8$&$-1.57$&$< 4$&$< 0.90$&$-0.50$ \\
$N(2120,3/2^-)$
&$4.0 \pm 0.2$&$-1.91$&$ $&$ $&$-1.91$ \\
\hline
$N(1685,1/2^+)$
& & & & &$1.4 $ \\
\hline
\end{tabular}
\end{table}

We want to mention that the experimental data on the nucleon
resonances  in the 2012 edition of the PDG~\cite{Beringer:1900zz}  
were much changed from those in the 2010 edition~\cite{Nakamura:2010zzi}.  
The $J^P = 5/2^+$ state $F_{15}(2000)$ is split into $N(1860,5/2^+)$ and 
$N(2000,5/2^+)$. The $D_{13}(2080)$ is split into $N(1875,3/2^-)$ and
$N(2120,3/2^-)$. The $S_{11}(2090)$ is changed into $N(1895,1/2^-)$ and the 
$N(2060,5/2^-)$ was previously known as $D_{15}(2200)$. 
The quark model predictions for the decay
amplitudes~\cite{Capstick:1998uh} are obtained from the nucleon
resonances before the 2012 edition of the PDG. Thus we continue to
compute the observables of $\eta$ photoproduction off the neutron on
the assumption that the quark model can reliably produce the values of
the decay amplitudes. 

For the $s$-channel diagrams with the higher-spin nucleon resonances,
we introduce the gaussian form factor that can suppress sufficiently
the cross sections when the energy
grows~\cite{Corthals:2005ce,DeCruz:2012bv} 
\begin{align}
F_{\mathrm{N^*}}(s) = \mathrm{exp}
\left\{ - \frac{(s-M_{N^*}^2)^2}{N_{N^*}^4} \right\}.
\label{eq:FF2}
\end{align}
Note that the phase factors for the nucleon resonance cannot be
determined just by symmetries, so that we will choose them
phenomenologically. The corresponding invariant amplitudes
are written by
\begin{align}
\mathcal{M}_{\mathrm{Res}} =
\sum_{N^*} e^{i\psi_{N^*}} {\mathcal M_{N^*}} F_{N^*}(s).
\label{eq:ResAmp}
\end{align}
We use the the values of the cutoff mass $\Lambda_{N^*} = 1.0$ GeV
again to avoid any additional ambiguities. The phase factor is chosen
to be $e^{i\psi_{N^*}} = e^{i\pi/2}$.

\section{Results and discussions}
\label{SecIII}
\begin{figure}[htp]
\includegraphics[width=8.3cm]{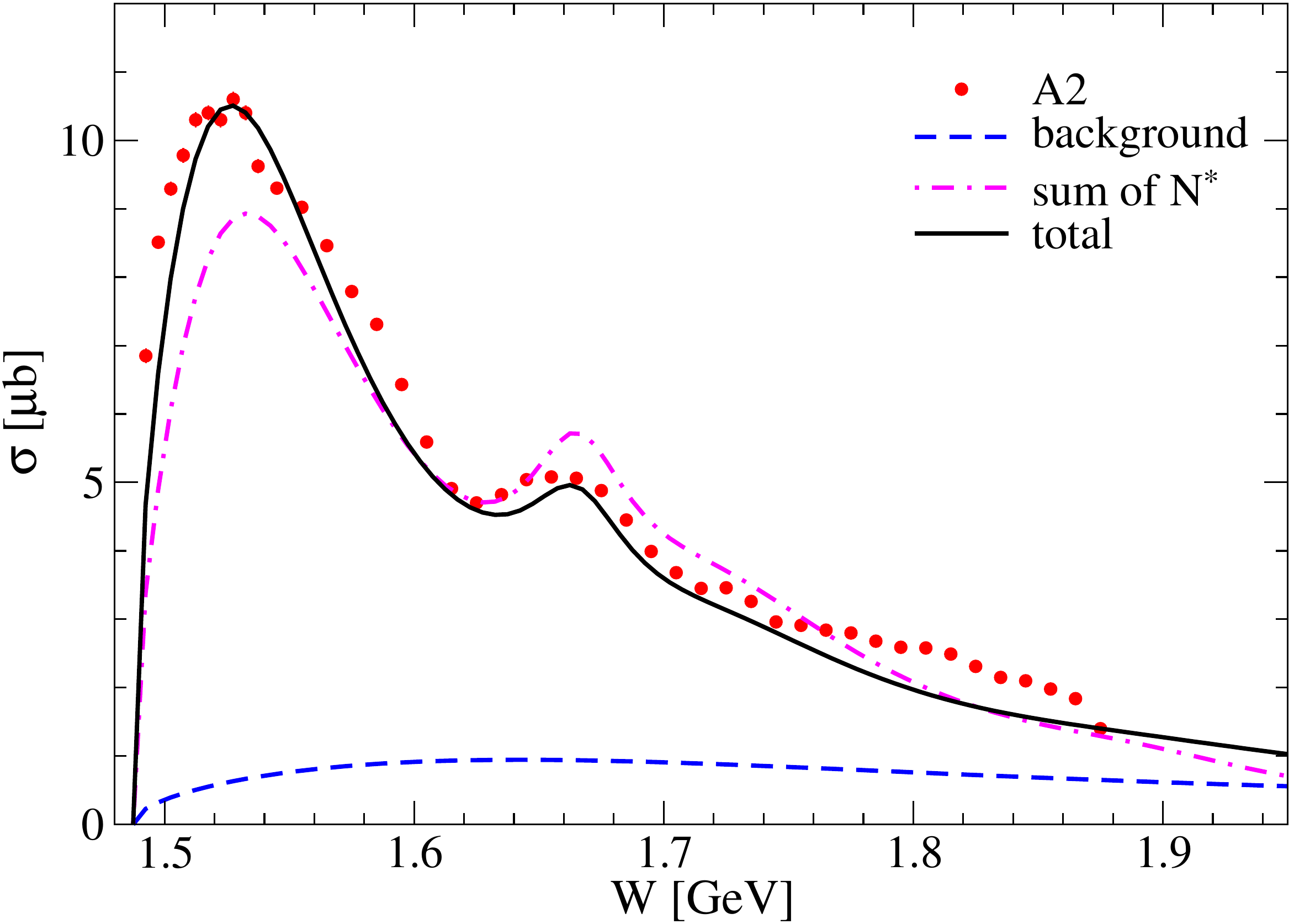} \,\,\,
\includegraphics[width=8.3cm]{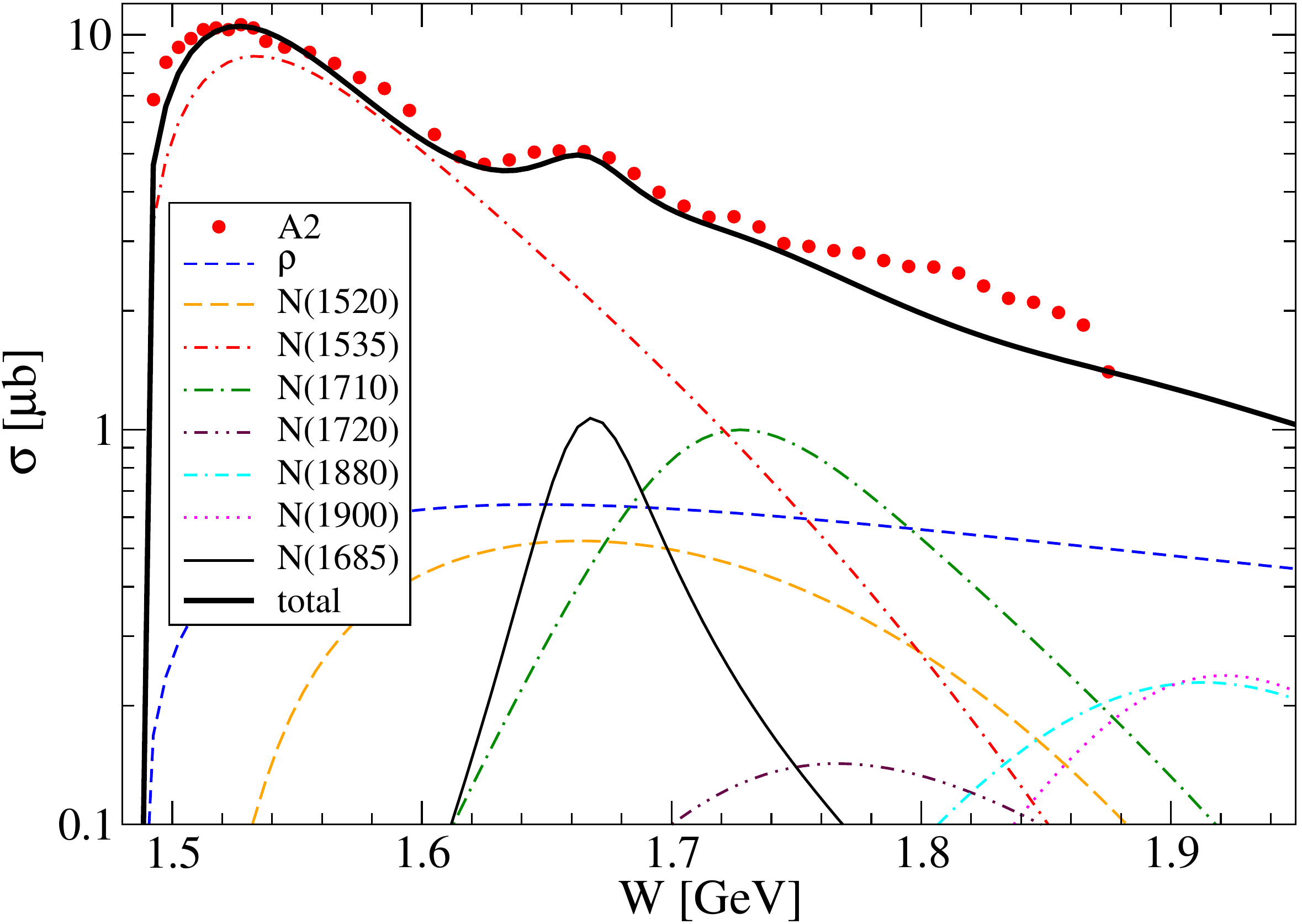}
\caption{Total cross section for the $\gamma n \to \eta n$ reaction
as a funciton of $W$.
The left panel draws the numerical result of the present work. 
The dashed curve represents the background contributions from the nucleon,
$\omega$-, and $\rho$-meson trajectories, whereas the dot-dashed curve 
draws the contribution of the nucleon resonances.
The solid curve plots the total contribution.
The right panel depicts each contribution of the exchanged particles to 
the total cross section for the $\gamma n \to \eta n$. 
The legend in the box indicates each contribution in a different colored 
curve.
The experimental data are taken from the A2 Collaboration
~\cite{Werthmuller:2014thb}.} 
\label{fig:2}
\end{figure}

We are now in a position to disscuss the present numerical results for
the $\gamma n \to \eta n$ reaction. In the left panel of
Fig.~\ref{fig:2}, the total cross section is plotted as a function of
the CM energy $W$. Though we do not display
explicitly the contribution of Reggeon-exchanges, $\rho$-Reggeon
exchange is in fact the strongest one among the background contributions 
($N$, $\rho$-, and $\omega$-Reggeon exchanges). 
This can be easily understood from the fact that both the $\rho$-meson
radiative and strong coupling constants are larger than those of the
nucleon and $\omega$ meson (see Eqs.~(\ref{eq:Coupl:VEtaG}) and
(\ref{eq:Coupl:NP})). However, the effects of the background
contributions turn out to be rather small on the total cross
section. Even at relatively high energies (1.7 GeV $\leq$ W $\leq$ 1.9
GeV), the magnitudes of the background contributions reach only the
level of around $30\,\%$ compared to the total result. 
On the other hand, the nucleon resonances play crucial roles of
describing the total cross section through the whole  
energy region from threshold up to $W = 1.9$ GeV. 
The present result is in good agreement with the data of the A2
Collaboration~\cite{Werthmuller:2014thb}.

The right panel of Fig.~\ref{fig:2} draws each contribution of the
$\rho$-Reggeon exchange and the excited nucleon resonances to the
total cross section in the logarithmic scale. Note that the
contributions of the only seven nucleon resonances are presented,
since all other nucleon resonances contribute almost negligibly  
to the total cross section. As explained in the previous Section, we
consider a total of fifteen nucleon resonances from the
PDG~\cite{Tanabashi:2018} and in addition the narrow resonance
$N(1685,1/2^+)$. It turns out that $N(1535,1/2^-)$ is predominantly
responsible for the description of the total cross section and
$N(1520,3/2^-)$ and $N(1710,1/2^+)$ also have sizable effects on it. 
On the other hand, $N(1720,3/2^+)$, $N(1880,1/2^+)$, and $N(1900,3/2^+)$ 
make rather small contributions to the results of the total cross section. 
It is very interesting to see that the narrow resonance $N(1685,1/2^+)$ 
indeed describes the narrow bump-like structure around $W\sim 1.68$ GeV. 
Though we have not shown the contributions of other nucleon resonances
that affect the total cross section negligibly, we will examine their
effects on polarization observables later. It is worth to compare our
results with other model results. The new version of the EtaMAID
model~\cite{Kashevarov:2018vpj} involves the analysis of the $\gamma n
\to \eta n$ reaction and reaches a different conclusion from the
present one. It is demonstrated that besides the dominant
$N(1535,1/2^-)$,  $N(1700,3/2^-)$, $N(1710,1/2^+)$, $N(1720,3/2^+)$,
$N(1880,1/2^+)$, and  $N(1895,1/2^-)$ make important contributions to
the cross sections. That is, the narrow structure is explained without
introducing $N(1685,1/2^+)$ but by the interference between other
$s$, $p$, and $d$ waves. This interpretation is also distinguished
from Ref.~\cite{Anisovich:2017afs}. 

\begin{figure}[hbp]
\centering
\includegraphics[width=17cm]{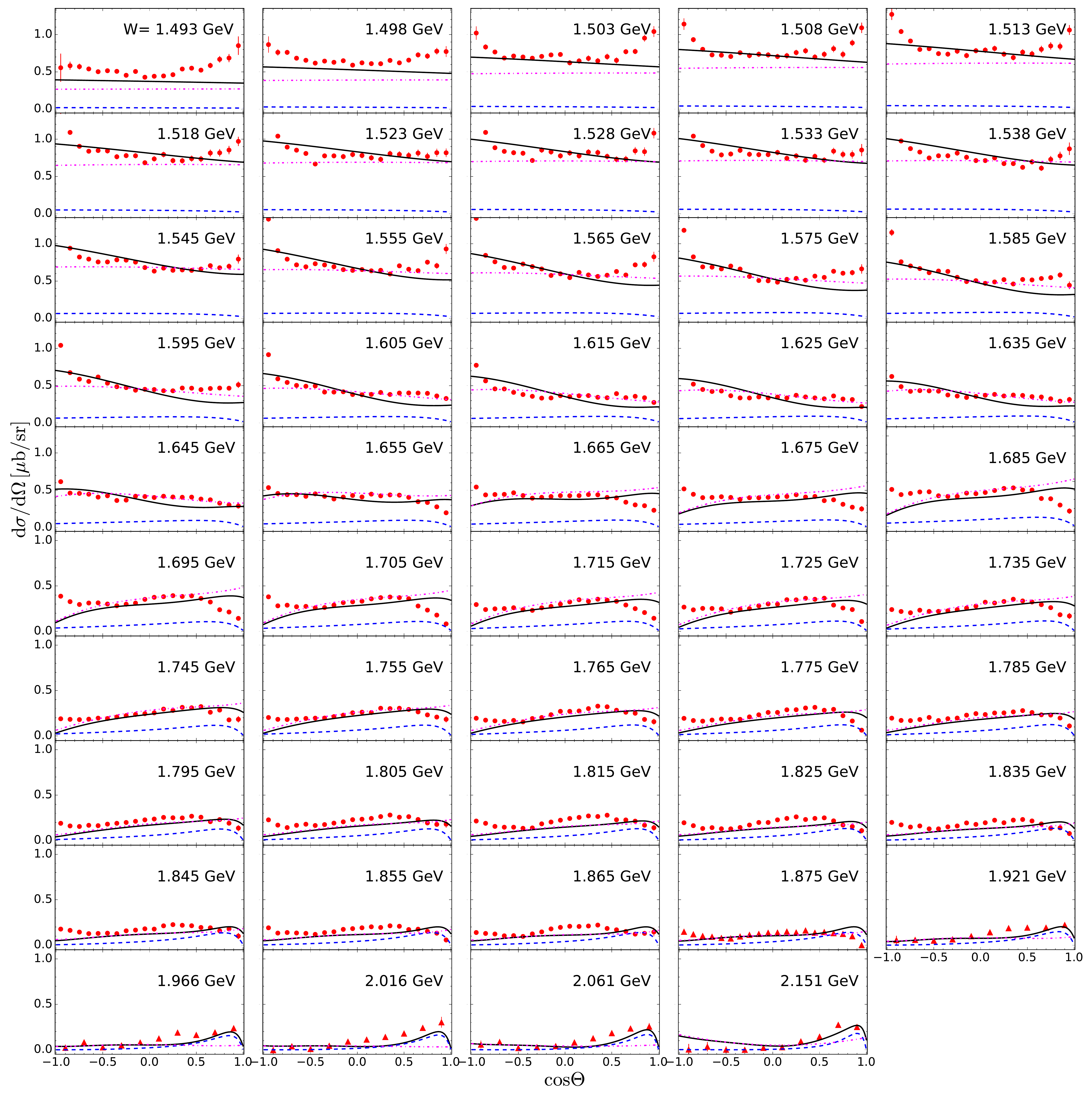}
\caption{Differential cross sections for the $\gamma n \to \eta n$ as a 
function of $\cos\theta$ for each beam energy.
The notations are the same as in the left panel of Fig.~\ref{fig:2}.
The data are from the A2 Collaboration (circle)~\cite{Werthmuller:2014thb} 
and CBELSA/TAPS Collaboration (triangle)~\cite{Jaegle:2008ux}.}
\label{fig:3}
\end{figure}
The value of the coupling constant $g_{\eta N N(1685,1/2^+)}$ is fitted to 
be 1.4 in our treatment of $\eta n$ photoproduction to reproduce the A2 
data~\cite{Werthmuller:2013rba}.
It corresponds to the branching ratio Br($N(1685,1/2^+) \to \eta N$) = 
$5.5\,\%$ with $\Gamma_{N(1685,1/2^+)} = 30$ MeV. 
In the meanwhile, the following upper limit is extracted from the A2 
data~\cite{Werthmuller:2014thb,Werthmuller:2013rba}
\begin{align}
\sqrt{\mathrm{Br}(N(1685) \to \eta N)} \, A_{1/2}^n <
(12.3 \pm 0.8) \times 10^{-3} \, \mathrm{GeV}^{-1/2},
\label{eq:BrUppLim}
\end{align}
which is quite close to the present result, i.e., 
$12.1 \, \times 10^{-3} 
\,\mathrm{GeV}^{-1/2}$. It is also of great interest to compute
the relative branching ratios of $\eta N$ and $K\Lambda$ based on the
present model. Together with results from a recent work on
$K^0\Lambda$ photoproduction~\cite{Kim:2018qfu}, we obtain 
\begin{align}
\frac{\mathrm{Br}(N(1685) \to K \Lambda)}
{\mathrm{Br}(N(1685) \to \eta N)} \simeq 0.15,
\label{eq:BrUppLim2}
\end{align}
which is consistent with the yield by a Bn-Ga partial wave analysis,
the same experimental data being used~\cite{Werthmuller:2014thb,
Werthmuller:2013rba,Anisovich:2017afs}:
\begin{align}
\frac{\mathrm{Br}(N(1685) \to K \Lambda)}
{\mathrm{Br}(N(1685) \to \eta N)} < \frac{1}{4}.
\label{eq:BrUppLim-2}
\end{align}

In Fig.~\ref{fig:3}, the differential cross sections are plotted as a 
function of $\cos\theta$ in the range of $W = 1.5\, \mathrm{GeV} - 2.1\,
\mathrm{GeV}$, where $\theta$ is the scattering angle of the 
$\eta$-meson in the CM frame. As expected from Fig.~\ref{fig:2}, $N^*$
contributions are the most important ones to describe the
A2~\cite{Werthmuller:2014thb} and CBELSA/TABS~\cite{Jaegle:2008ux}
data. However, the background contribution also helps to improve the
experimental data, especially at higher energies.
The effect of the background is small near the threshold but gets larger 
as photon energy $W$ increases.

\begin{figure}[htp]
\centering
\includegraphics[width=8.3cm]{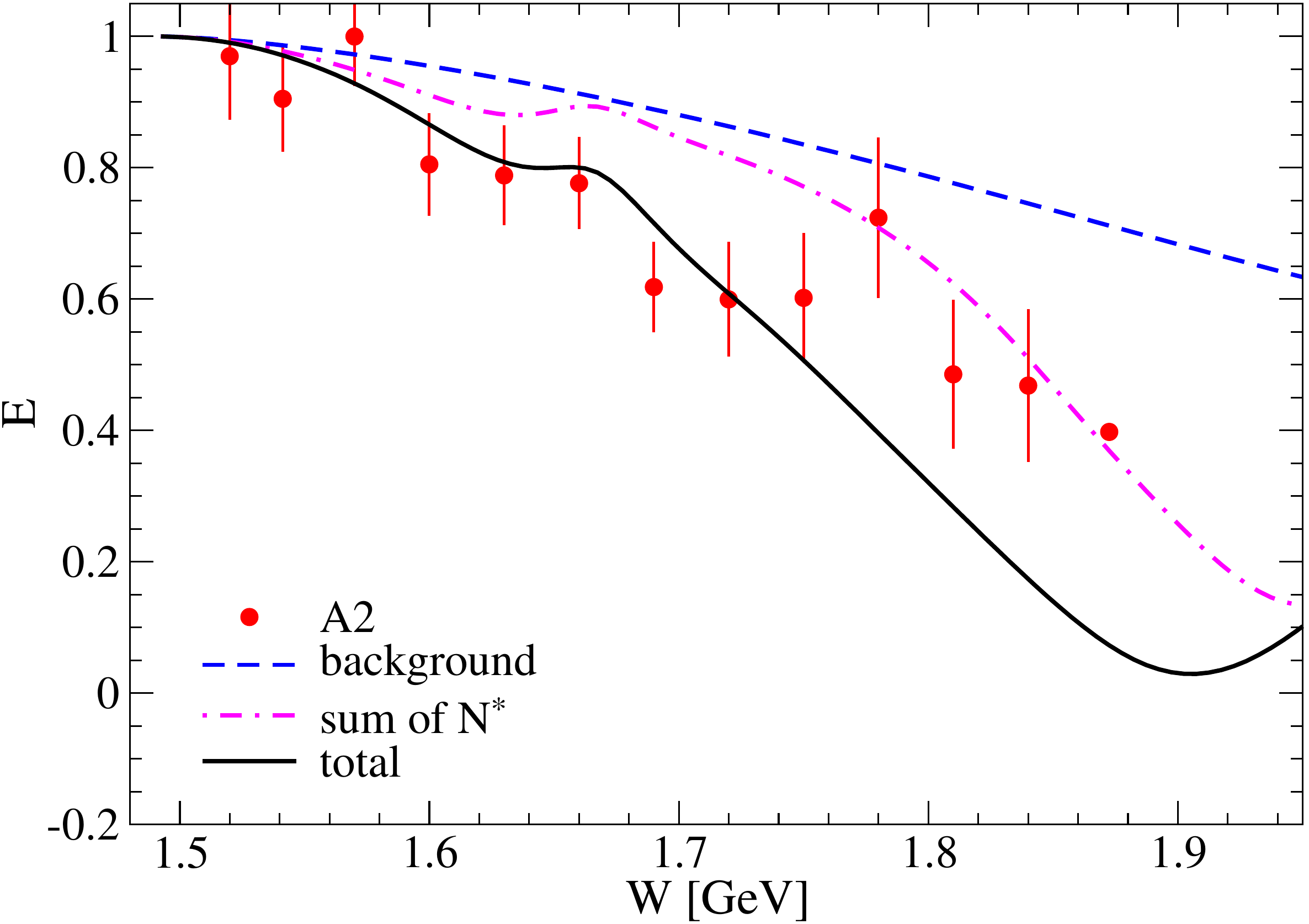} 
\caption{Double polarization observable $E$ for the $\gamma n \to \eta n$
as a function of $W$. The notations are the same as in the left panel
of Fig.~\ref{fig:2}. The data are taken from the A2
Collaboration~\cite{Witthauer:2017get}.} 
\label{fig:4}
\end{figure}
\begin{figure}[htp]
\includegraphics[width=8.3cm]{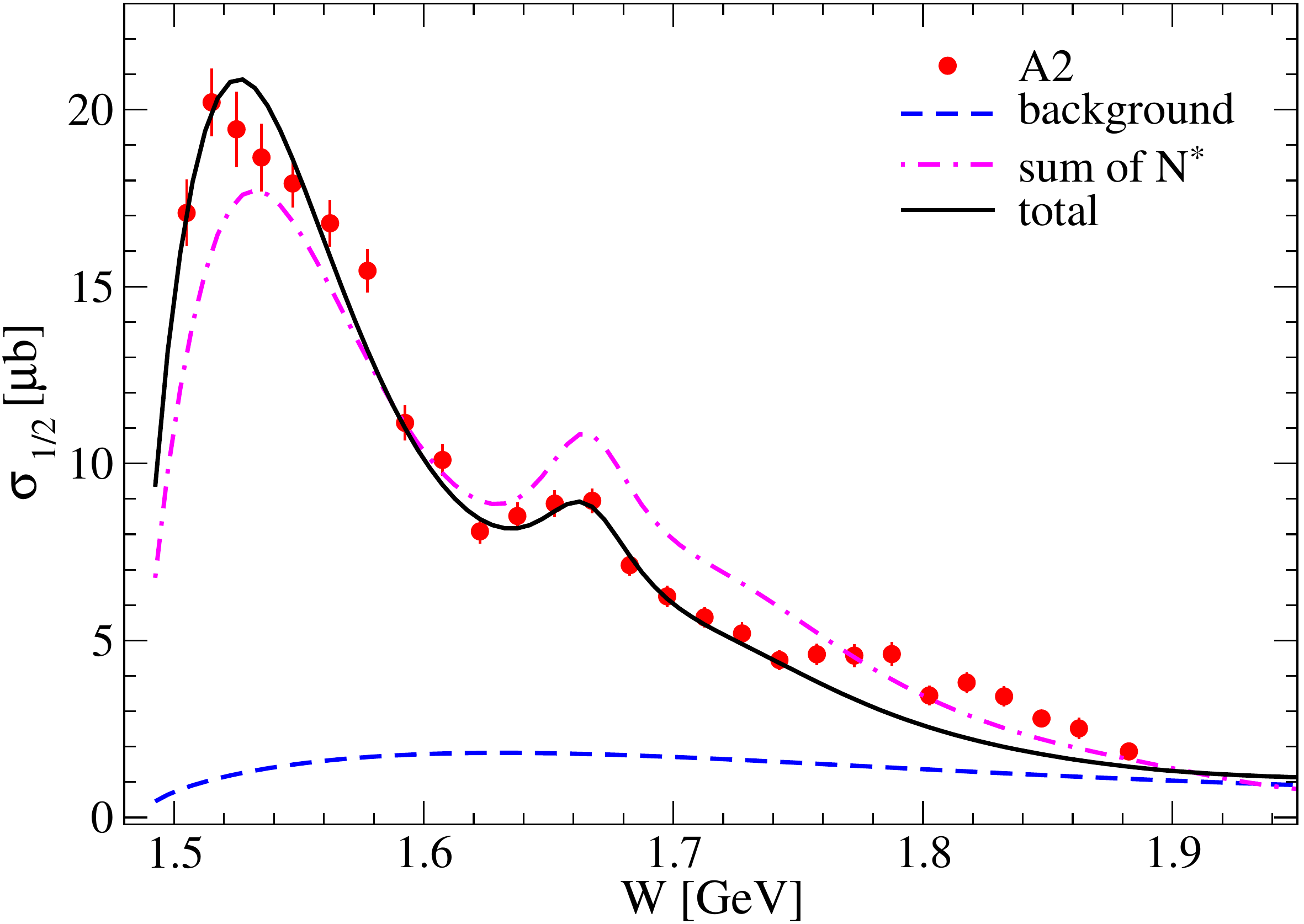} \,\,\,
\includegraphics[width=8.3cm]{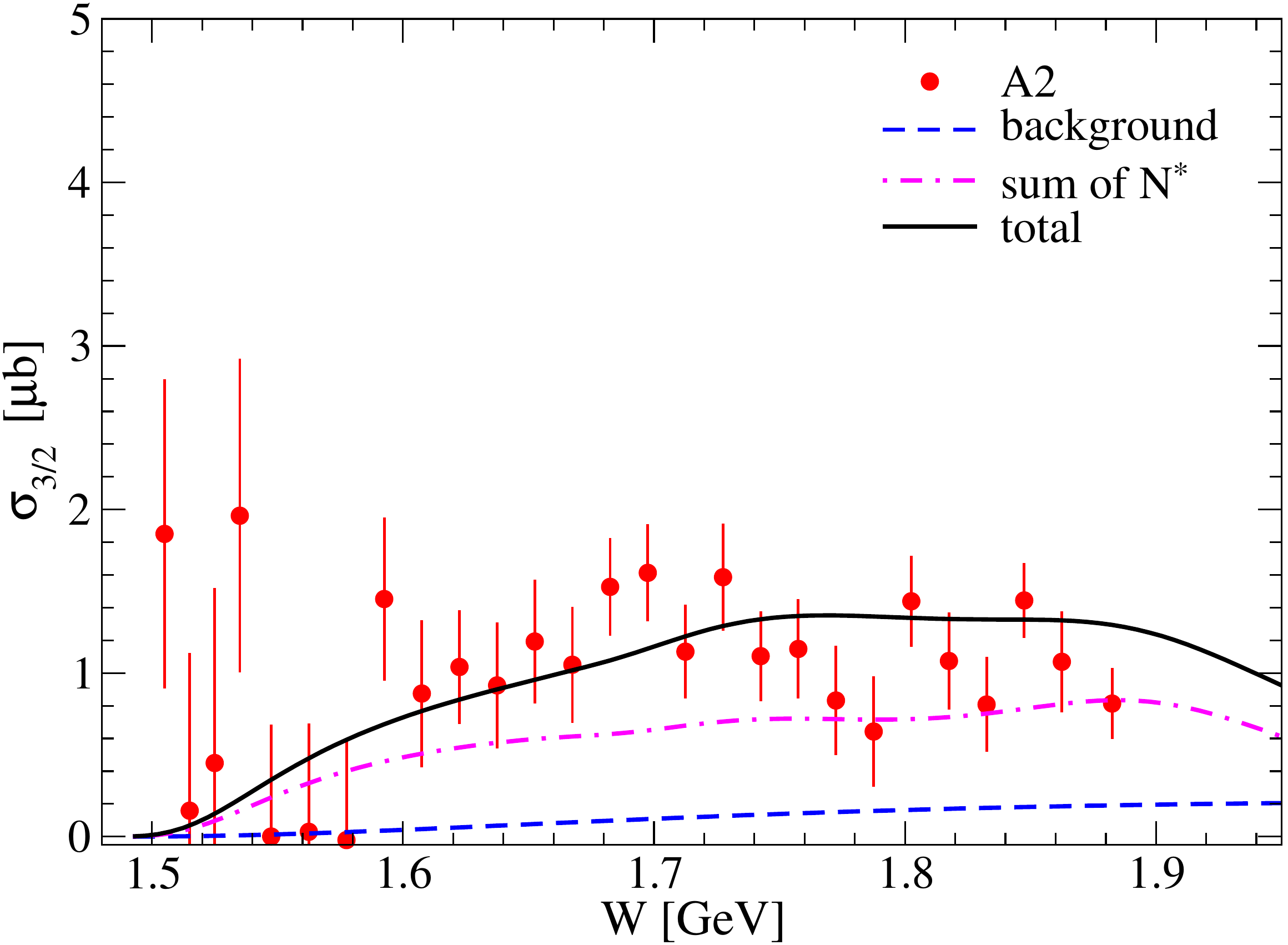}
\caption{Helicity-dependent total cross section for the 
$\gamma n \to \eta n$ as a function of $W$.
The notations are the same as in the left panel of Fig.~\ref{fig:2}.
The data are taken from the A2 Collaboration~\cite{Witthauer:2017get}.}
\label{fig:5}
\end{figure}
Since a number of nucleon resonances are involved in the present model, 
it is of importance to compute various polarization observables to pin 
down the role of each nucleon resonance more properly.
In Fig.~\ref{fig:4}, the double polarization observable $E$ is drawn as a
function of $W$ and compared with the A2 data~\cite{Witthauer:2017get}.
It is defined by~\cite{Witthauer:2017get,Barker:1975bp,Fasano:1992es}
\begin{align}
E =
\frac{\sigma^{(r,+z,0)}-\sigma^{(r,-z,0)}}{\sigma^{(r,+z,0)}+\sigma^{(r,-z,0)}}
= \frac{\sigma_{1/2}-\sigma_{3/2}}{\sigma_{1/2}+\sigma_{3/2}},
\label{eq:DP:E}
\end{align}
where the superscript $(B,\,T,\,R)$ stands for the polarization state of
the photon beam, target nucleon, and recoiled nucleon, respectively.
We assume that the reaction takes place in the $x - z$ plane.
Thus the $+z (-z)$ direction indicates the incident beam and
longitudinally polarized target is parallel (antiparallel) to each other.
$r$ designates the circularly polarized photon beam with helicity $+1$.
The result of the beam-target asymmetry $E$ only from the background 
contribution decreases gradually as the photon energy $W$ increases and 
lies above the A2 data. The inclusion of the nucleon resonances 
pulls down the beam-target asymmetry and it finally reaches zero at 
$W=1.9$ revealing some bump structure near $W = 1.68$ GeV.

Figure~\ref{fig:5} plots the helicity-dependent cross sections
$\sigma_{1/2}$ and $\sigma_{3/2}$ extracted from the beam-target
asymmetry $E$~\cite{Witthauer:2017get} and the unpolarized cross
section $\sigma_0$ ~\cite{Werthmuller:2014thb}, defined by
\begin{align}
\sigma_{1/2}=\sigma_0(1+E), \,\,\, \sigma_{3/2}=\sigma_0(1-E).
\label{eq:HeliDepCS}
\end{align}
The left panel of Fig.~\ref{fig:5} clearly shows that the nucleon 
resonances with spin $J=1/2$ is correctly described.
The total result is in excellent agreement with the A2 data.
More specifically, the background term makes a constructive interference 
effect with $N(1535,1/2^-)$ and a destructive effect with $N(1685,1/2^+)$
and $N(1710,1/2^+)$. The result of $\sigma_{3/2}$ is also in good
agreement with the A2 data. The constructive interference effect
between the  backgound and resonance contributions is rather important
as shown in the right panel of Fig.~\ref{fig:5}.
It indicates that the nucleon resonances with higher spin $J \geq 3/2$
are properly considered in the present framework. Thus the nucleon
resonances with $J=3/2$, i.e., $N(1520,3/2^-)$, $N(1720,3/2^+)$, and
$N(1900,3/2^+)$ come into play for the description of the $\gamma
n \to \eta n$ in addition to the dominant spin-1/2 nucleon resonances
$N(1535,1/2^-)$, $N(1685,1/2^+)$, and  $N(1710,1/2^+)$.
The results of Fig.~\ref{fig:5} imply that the photo- and
strong-coupling constants are well constrained by this model 
calculations (see Tables~\ref{TAB1} and~\ref{TAB2}).

\begin{figure}[h]
\includegraphics[width=8.3cm]{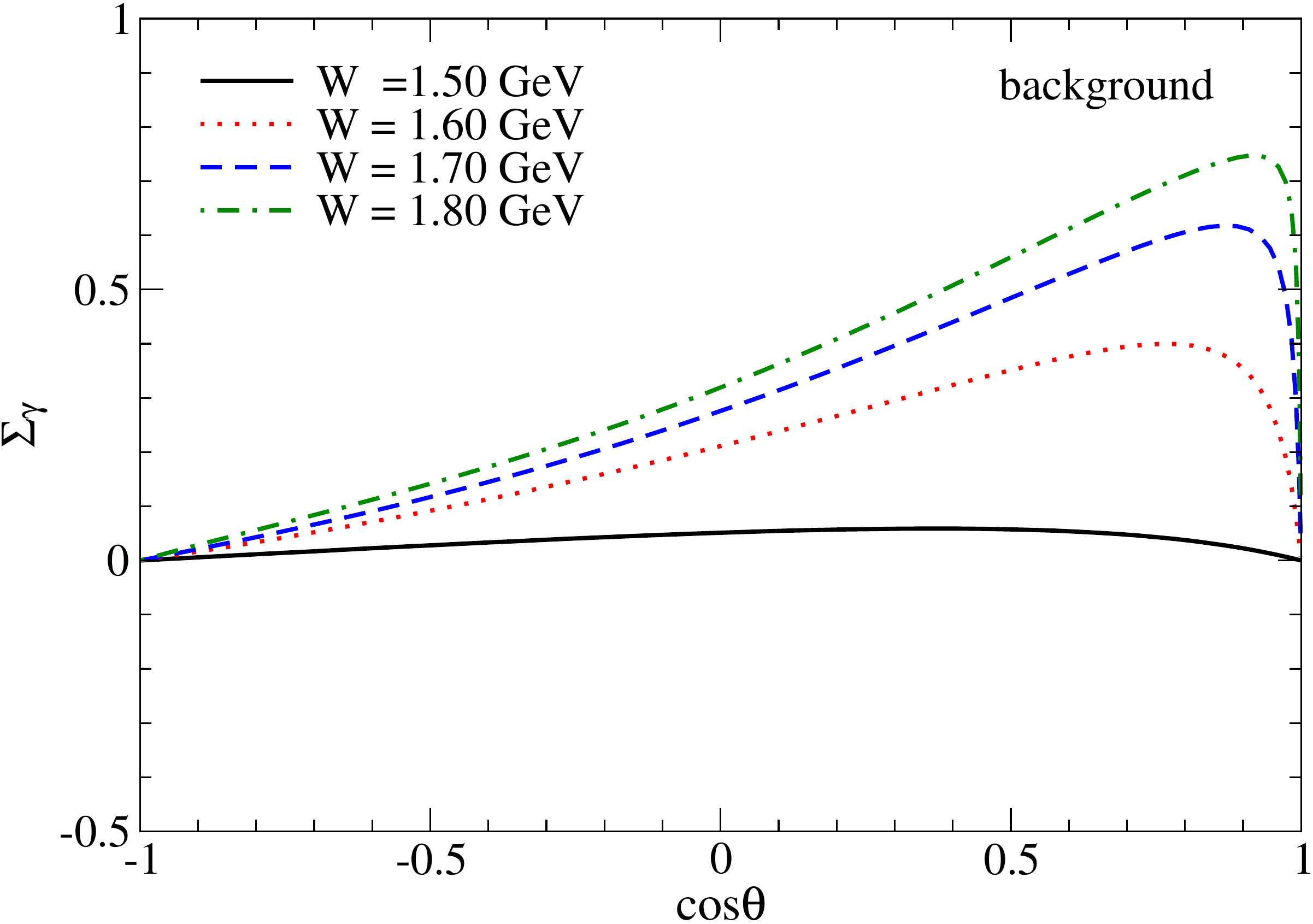} \,\,\,
\includegraphics[width=8.3cm]{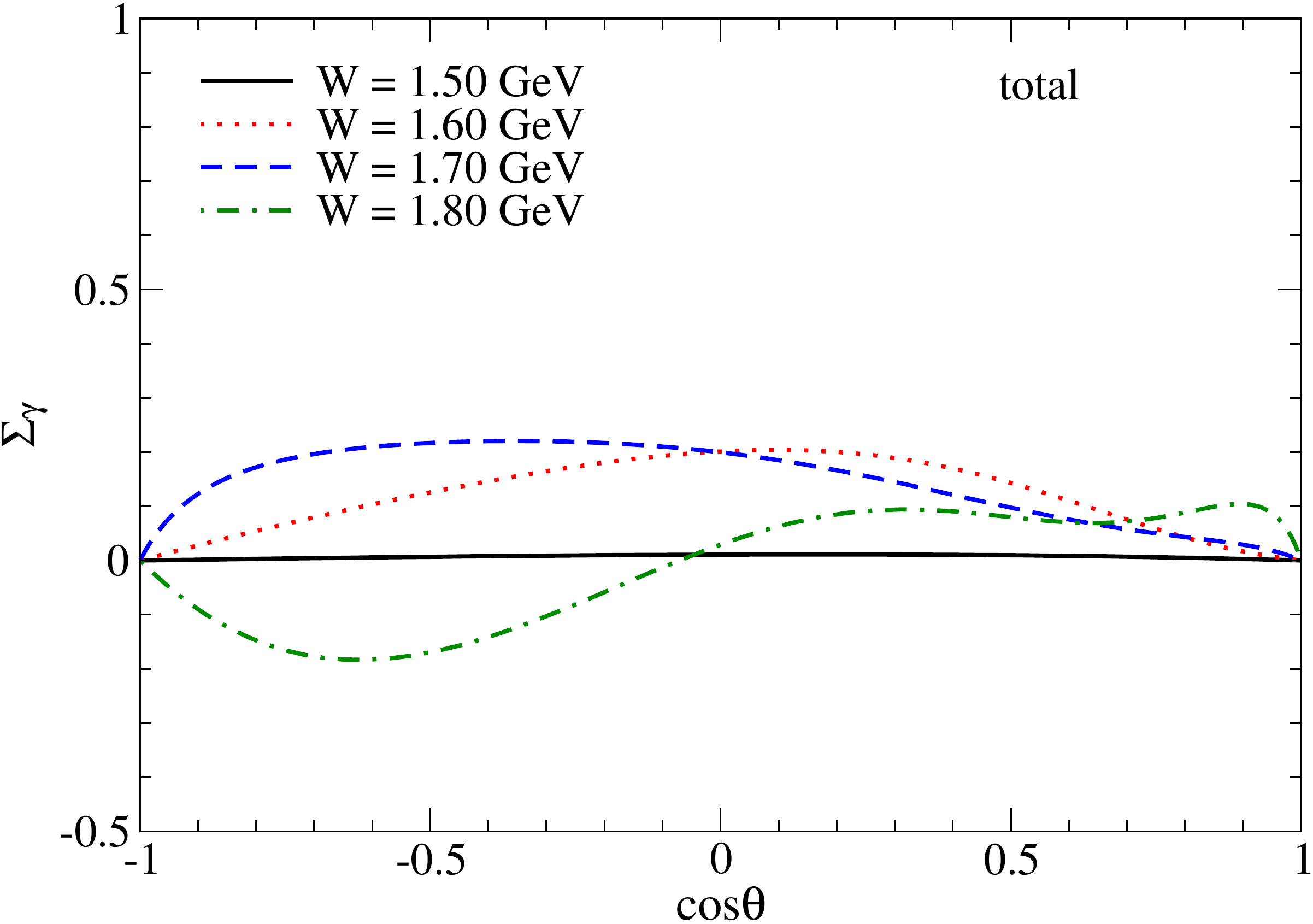}
\caption{Beam asymmetry for the $\gamma n \to \eta n$ as a function of 
$\cos\theta$.}
\label{fig:6}
\end{figure}
In Fig.~\ref{fig:6}, the predictions of the beam asymmetry
$\Sigma_{\vec{\gamma} n \to \eta n}$ are drawn as a function of
$\cos\theta$ for four different photon energies.
It is defined by
\begin{equation}
\label{eq:BA}
\Sigma_{\vec{\gamma} n \to \eta n} = 
\frac{\frac{d\sigma}{d\Omega}_\perp-\frac{d\sigma}{d\Omega}_\parallel}
{\frac{d\sigma}{d\Omega}_\perp+\frac{d\sigma}{d\Omega}_\parallel},
\end{equation}
where the subscript $\perp$ means that the photon polarization vector
is perpendicular to the reaction plane whereas $\parallel$ stands for the
parallel photon polarization to it.
The left panel of Fig.~\ref{fig:6} depicts the results of the beam 
asymmetry only from the background contribution.
Because of the dominant vector-meson Regge trajectories, the curves
gradually increases as $\cos\theta$ increases and then falls off
drastically at very forward angles. When the nucleon resonances are
included as shown in the right panel of Fig.~\ref{fig:6}, the changes
are dramatic. The beam asymmetry gets diminished and the
magnitudes of the $\Sigma_{\vec{\gamma} n \to \eta n}$ becomes overall
equal or less than 0.2. The measurements of the $\Sigma_{\vec{\gamma}
  n \to   \eta n}$ as well as other polarization observables by future
experiments will become a touchstone to judge which interpretation
will turn out right.  

\section{Summary and conclusion}
\label{SecIV}
In the present work, we investigated $\eta n$ photoproduction off the
neutron taking into account fifteen nucleon resonances from
the PDG and in addition the narrow nucleon resonance $N(1685,1/2^+)$
in the $s$-channel diagram. We employed an effective Lagrangian approach
combining with a Regge model. The $t$-channel $\rho$ and $\omega$
Regge trajectories and $N$ exchange in the $s$ channel are
considered as a background.
The photo- and strong-coupling constants for the resonance terms were
all fixed within the range of values taken from the experimental data
and quark model predictions without any complex fitting procedure.
We were able to reproduce quantitatively the A2 data for the total and
differential cross sections.   
The branching ratio of the $N(1685,1/2^+)$ to the $\eta N$ channel and
the relative branching ratio of it to the $K \Lambda$ channel were also 
studied. It turned out that the branching ratio of the $N(1685,1/2^+)$
in the  $\eta N$ channel is almost ten times larger than that to the
$K  \Lambda$ channel. 

The double polarization observables $E$ and the related helicity-dependent
cross sections $\sigma_{1/2}$ and $\sigma_{3/2}$ were also examined, since
they are useful to clarify the role of spin $J=1/2$ and $J \geq 3/2$
nucleon resonances separately. We have found that $N(1685,1/2^+)$ and
$N(1710,1/2^+)$ are the dominant contributions to the $\gamma n \to
\eta n$ apart from the $N(1535,1/2^-)$ that is the most dominant one.
Other nucleon resonances such as $N(1520,3/2^-)$, $N(1720,3/2^+)$, 
$N(1880,1/2^+)$, and $N(1900,3/2^+)$ also come into play in describing
the A2 data.

\begin{acknowledgments}
The authors thank D. Werthm{\"u}ller for providing the A2 experimental
data. H.-Ch. K. is grateful to A. Hosaka, T. Maruyama,
M. Oka for useful discussions. He wants to express his gratitude to
the members of the Advanced Science Research Center at Japan Atomic
Energy Agency for the hospitality, where part of the present work was
done. This work was supported by the National Research Foundation of
Korea (NRF) grant funded by the Korea government(MSIT)
(No. 2018R1A5A1025563).  
\end{acknowledgments}


\end{document}